\documentstyle[12pt,epsf]{article}

\def\be{\begin{equation}}
\def\ee{\end{equation}}
\def\bea{\begin{eqnarray}}
\def\eea{\end{eqnarray}}

\begin{document}

\begin{flushright}
hep-th/0108198
\end{flushright}

\pagestyle{plain}

\def\e{{\rm e}}
\def\haf{{\frac{1}{2}}}
\def\tr{{\rm Tr\;}}
\def\goes{\rightarrow}
\def\ie{{\it i.e.}, }
\def\tcl{T_{\rm cl}}
\def\Goes {\Rightarrow}
\def\CX{{\cal X}}
\def\cm{{\rm c.m.}}
\def\fp{{\rm f.p.}}
\def\ca{{\cal A}}
\def\cf{{\cal F}}
\def\cd{{\cal D}}
\def\cv{{\cal V}}
\def\cvsym{{\cal V}_{{\rm sym.}}}
\def\cvnonsym{{\cal V}_{{\rm non-sym.}}}
\def\iphi{{\bf i}_\Phi}
\def\iv{{\bf i_v}} 
\def\xgphi{{x\goes\Phi}}

\begin{center}
\vspace{2cm}
{\Large {\bf On The Interaction Of D0-Brane Bound States}}\\
\vspace{.4cm}

{\Large {\bf And RR Photons}}
\vspace{.8cm}

Amir H. Fatollahi 

\vspace{.4cm}

{\it Dipartimento di Fisica, Universita di Roma ``Tor Vergata",}\\ {\it
INFN-Sezione di Roma II, Via della Ricerca Scientifica, 1,}\\ {\it 00133,
Roma, Italy}
\vspace{.2cm}

{\it and}
\vspace{.2cm}

{\it Institute for Advanced Studies in Basic Sciences (IASBS),}\\ {\it
P.O.Box 45195-159, Zanjan, Iran}
\footnote{Postal and permanent address.}

\vspace{.4cm}

{\sl fatho@roma2.infn.it}
\end{center}
\vskip .3 cm

\begin{abstract}
We consider the problem of the interaction between D0-brane bound state
and 1-form RR photons by the world-line theory. Based on the fact that in
the world-line theory the RR gauge fields depend on the matrix coordinates
of D0-branes, the gauge fields also appear as matrices in the formulation. 
At the classical level, we derive the Lorentz-like equations of motion for
D0-branes, and it is observed that the center-of-mass is colourless with
respect to the $SU(N)$ sector of the background. Using the path integral
method, the perturbation theory for the interaction between the bound
state and the RR background is developed. We discuss what kind of field
theory may be corresponded to the amplitudes which are calculated by the
perturbation expansion in world-line theory. Qualitative considerations
show that the possibility of existence of a map between the world-line
theory and the non-Abelian gauge theory is very considerable. 
\end{abstract}

\vspace{1.0cm}



\newpage

\section{Introduction}
\setcounter{equation}{0}
In recent years a great attention is appeared in formulation and studying
field theories on non-commutative spaces. Apart from the abstract
mathematical interests, the physical motivation in doing so has been the
natural appearances of non-commutative spaces in String Theory. 
Correspondingly, it has been understood that String Theory is involved by
some kinds of non-commutativities; two important examples are, 1) the
coordinates of the bound states of $N$ D$p$-branes \cite{9510017} are
presented by $N\times N$ hermitian matrices \cite{9510135}, and 2) the
longitudinal directions of D$p$-branes in the presence of NS B-field
background appear to be non-commutative \cite{CDS,9908142}, as are seen by
the ends of open strings \cite{jabbari}. In the second example, the
coordinates in the longitudinal directions of D$p$-branes act as some
operators and satisfy the algebra: 
\bea\label{algebra}
[\hat x^\mu,\hat x^\nu]=i\theta^{\mu\nu},
\eea
where $\theta^{\mu\nu}$ is a constant anti-symmetric tensor. There are a
lot of attempts in the recent literature to study different aspects of the
field theories defined on these kinds of non-commutative spaces. As a
point, we mention that the above algebra is satisfied just by $\infty
\times \infty$ dimensional matrices, and consequently, the concerned
non-commutativities should be assumed in all (regions) of the space. Also,
since there is a non-zero expectation value for a tensor field as $\langle
B^{\mu\nu} \rangle = (\theta^{-1})^{\mu\nu}$ \cite{9908142}, in these
spaces generally one should expect the violation of the Lorentz
invariance. 

As we recalled in above, there is another kind of non-commutativity
concerning the coordinates of D-brane bound states, calling them from now
on ``matrix coordinates". In contrast to the case related to the algebra
(\ref{algebra}), for the case of D-brane bound states, we have
non-commutativity for finite dimensional matrices, and thus the
non-commutativity of coordinates is not extended to all of the space. In
this case the non-commutativity is `confined' just inside the bound state;
saying by simple words: ``The non-commutativity is not seen by an observer
far from the bound state". In contrast to the case of infinite extension
of non-commutativity, we call this kind {\it confined non-commutativity}. 

By this picture, the natural question is: How can we know about the
structure of confined non-commutativity? Since the non-commutativity of
bound state is confined, like any other similar situation known in
physics, the answer to the above question is gained by analysing and
studying the response of the substructure of bound state to the external
probes. In this respect one may consider two kinds of the external probes,
1) another D-brane, or 2) quanta of external fields, like gravitons or
photons of form fields. To be specific, let us consider things in the
special case of D0-branes. Using another D0-brane, as a probe of a system
of D0-branes, is a familiar example from the studies related to the
M(atrix) model conjecture of M-theory \cite{9610043}. In the M(atrix)
model picture, since D0-branes are already assumed to be super-gravitons
of the 11-dimensional super-gravity theory in the light-cone gauge, the
problem in hand is in fact nothing but `probing' the bound state by
another individual `graviton'. There in M(atrix) model, the high amount of
supersymmetry, together with the specific form of the commutator potential
of matrix coordinates, help to calculate the elements of S-matrix for
various scattering processes. The important peculiarity of this case is
that, by these kinds of investigations one uses non-commutativity (by
things like the commutator potential) to study the effective theory of
D0-branes, rather than analysing the `structure' of confined
non-commutativity itself \cite{review-matrix}. In other words, generally
in this case one ignores the internal dynamics inside the bound state (as
target), and essentially considers only the relative dynamics of the
target and other D0-brane(s) as probe(s). 

In this work we want to discuss the basic elements of using the second
kind of the probes mentioned in above (\ie external fields) to find
information about the structure of confined non-commutativity. As it will
be clear through the paper, the language which is used by this kind of
probe is much closer to the field theory formulation of the problem, in
comparison with the approach in which the probe is viewed as another
D0-brane. To do this, we need to know the dynamics of the bound state of
D0-branes in different backgrounds. Due to the nature of matrix
coordinates, the formulation of the dynamics of D0-branes in the
background of gravity and various form fields is a nontrivial question.
One of the most important progresses in this direction is done by the
works \cite{9910053,9910052}. Here we use the results of
\cite{9910053,9910052}, restricting ourselves to the simplest case of zero
NS B-field and flat metric, but non-zero 1-form RR field. Though the
framework here we use is coming from D$p$-branes of String Theory, it is
useful to consider the more general case in arbitrary space-time
dimensions $d+1$. Also as the first step, we consider the bosonic partners
only.

One of the questions which can be addressed in this direction is about the
nature of the effective field theory which captures the interaction
between the bound state of D0-branes and the `photons' of the 1-form RR
field. To be more specific, it will be interesting to derive the effective
vertex function for the interaction of a 1-form RR photon with the
incoming--outgoing D0-brane bound states. These kinds of questions consist
some parts of the discussions of this paper. Also we discuss that the
amplitudes of which field theory can be corresponded to the amplitudes
which are derived by the world-line formulation of D0-branes in RR
background \footnote{The reader can refer to \cite{9907146}, as an attempt
to interpret the quantized propagation of D0-branes, while they are
interacting with each other via the commutator potential, as the Feynman
graphs of a field theory in the light-cone gauge.}. 

The world-line formulation we will use in this work, is very much like
that of M(atrix) model conjecture; in particular, it is in the
non-relativistic limit. To approach the Lorentz covariant formulation,
following finite-$N$ interpretation of \cite{9704080}, it is reasonable to
interpret things in the Discrete Light-Cone Quantization (DLCQ) framework. 
This point of view should be kept also for the correspondence we consider
to an effective field theory for the interacting theory of
D0-branes--photons. 

The organization of the remained parts of this paper is as follows.  In
Sec.2, based on \cite{9910053,9910052}, we review the main aspects of the
world-line formulation of the dynamics of D0-brane bound states in
nontrivial backgrounds. These include equations of motion of D0-branes in
1-form background, and also the symmetry aspects of the world-line
formulation. In Sec.3, by using the path integral method, we quantize the
D0-brane theory. In particular, we write down the expression of the
propagator in the first order of perturbation, which can be converted to
the amplitudes of the scattering processes by an arbitrary external
source. In Sec.4, we discuss the question of ``Which field theory can be
corresponded to the world-line theory of D0-branes in 1-form background?". 
Sec.5 is devoted to the conclusion and discussions. 

The discussions and ideas concerning in this paper were initiated by the
previous works of the author in \cite{0103262} and \cite{0104210}. In
particular, the problem we consider in this work interpreted in
\cite{0104210} as the world-line formulation of ``electrodynamics on
matrix space". Also the subject of `probing confined non-commutativity' is
mentioned briefly in the last part of \cite{0104210}. 

\section{On Dynamics Of D0-Branes In One-Form RR\\ Background}
\setcounter{equation}{0}
\subsection{First Look: D$p$-Branes In General Background}
It is known that the transverse coordinates of bound state of $N$
D$p$-branes, rather than numbers, are presented by $N\times N$ hermitian
matrices \cite{9510135}; see review \cite{9611050}. Due to the nature of
matrix coordinates, the formulation of the dynamics of D$p$-branes in the
background of gravity and various form fields is a nontrivial question.
One of the most important progresses in this direction is done by the
works \cite{9910053,9910052}. In \cite{9910053}, by taking T-duality of
String Theory as the guiding principle, an action for the dynamics of the
bound states of D$p$-branes in nontrivial background is proposed. The
proposed bosonic action for the bound state of $N$ D$p$-branes (in units
$2\pi l^2=1$) is the sum of:
\bea
\!\!\!\!\!\!\!\!\!\!\!
&~&S_{BI}=-T_p\int d^{p+1}\sigma\; \tr\bigg(\e^{-\phi}
\sqrt{-\det(P\{E_{IJ}+E_{Ii}(Q^{-1}-\delta)^{ij}E_{jJ}\}+F_{IJ})
\det(Q^i_j)}\bigg),\\
\!\!\!\!\!\!\!\!\!\!\!
&~&S_{CS}=\mu_p\int\tr\bigg(P\{\e^{i\;\iphi\iphi}(\sum
C^{(n)}\e^{B})\}
\e^{F}\bigg),
\eea
with the following definitions \cite{9910053}:
\bea
&~&E_{\mu\nu}\equiv G_{\mu\nu}+B_{\mu\nu},\;\;\;\;
Q^i_j\equiv \delta^i_j+i[\Phi^i,\Phi^j]E_{kj},\\
&~&\mu,\nu=0,\cdots,9,\;\;\;I,J=0,\cdots,p,\;\;\;i,j=p+1,\cdots,9.\nonumber
\eea
In the above $G_{\mu\nu}$ and $B_{\mu\nu}$ are the metric and NS B-field
respectively, and $\Phi^i$ are world-volume scalars and $N\times N$
hermitian matrices, that describe the position of D$p$-branes in the
transverse directions. The $C^{(n)}$ is $n$-form RR field, while $F_{IJ}$
is the $U(N)$ field strength. In this action, $P\{\cdots\}$ denotes the
pull-back of the bulk fields to the world-volume of the D$p$-branes, and
$\tr$is trace on the gauge group. $\iv$ denotes the interior product with
a vector ${\bf v}$; for example, $\iphi$ acts on the 2-form $C^{(2)}=\haf
C^{(2)}_{ij} dx^i dx^j$ as
\bea
\iphi C^{(2)}= \Phi^i C^{(2)}_{ij} dx^j,\;\;\;\;\;\;
\iphi\iphi C^{(2)}= \Phi^i \Phi^j C^{(2)}_{ij}=\haf [\Phi^i,\Phi^j]
C^{(2)}_{ij}.
\eea
Therefore $(\iphi)^2C^{(n)}$=0 for the commutative case, \ie for one
D$p$-brane. 

Some comments on the above action are in order:

{\it i)} All the derivatives in the longitudinal directions should be
actually covariant derivatives, \ie $\partial_I\goes D_I=\partial_I
+i[A_I,\;]$ \cite{hulldorn}. This point is true also for the pull-back
quantities. 

{\it ii)} The pull-back quantities depend on the transverse directions of
the D$p$-branes only via their functional dependence on the world-volume
scalars $\Phi^i$ \cite{douglas}. Since the matrix coordinates $\Phi$ do
not commute with each other, the problem of ordering ambiguity is present.
Following some arguments, it is proposed that the coordinates $\Phi$'s
appear in the background fields by the ``symmetrization prescription" 
\cite{9910053,garousi,9712185,9712159}. The symmetrization on
coordinates can be obtained by the so-called ``non-Abelian Taylor
expansion". The non-Abelian Taylor expansion for an arbitrary function
$f(\Phi^i,\sigma^I)$ is given by
\bea\label{NTE}
f(\Phi^i,\sigma^I)&\equiv&f(x^i,\sigma^I)|_\xgphi=\exp[\Phi^i\partial_{x^i}] 
f(x^i,\sigma^I)|_{x=0}\nonumber\\
&=&\sum_{n=0}^\infty
\frac{1}{n!} \Phi^{i_1}\cdots \Phi^{i_n} 
(\partial_{x^{i_1}}\cdots\partial_{x^{i_n}})
f(x^i,\sigma^I)|_{x=0}.
\eea
In the above expansion the symmetrization is recovered via the symmetric
property of the derivatives inside the term $(\partial_{x^{i_1}} \cdots
\partial_{x^{i_n}})$. 

{\it iii)} This action involves a single $\tr$, and this $\tr$should be
calculated by symmetrization prescription for the non-commutative
quantities $F_{IJ}$, $D_I\Phi^i$ and $i[\Phi^i, \Phi^j]$ \cite{tseytlin}
\footnote{There is a stronger prescription, with symmetrization between
all non-commutative objects $F_{IJ}$, $D_I\Phi^i$, $i[\Phi^i, \Phi^j]$,
and the individual $\Phi$'s appearing in the functional dependences of the
pull-back fields \cite{9910053,9904095}. We will not use this one in our
future discussions for the case of D0-branes, with no essential change in
the conclusions.}. 

To become more familiar with the terms in the action of D$p$-branes, let
us consider the special case $p=0$ of D0-branes, in which the world-volume
consists only the time direction, as $\sigma^0=t$. The dynamics of
D0-branes in background of metric $G_{\mu\nu}(x,t)$, the 1-form RR field
$C^{(1)}_\mu(x^\nu)\equiv A_\mu(x,t)$ and zero NS B-field, not being
precise about the indices and coefficients, in the lowest orders is given
by an action like \cite{9910053,9910052}: 
\bea\label{SD0-general} 
&S&=\int dt\; \tr \bigg(\frac{m}{2} G_{ij}(\Phi,t) D_t\Phi^iD_t \Phi^j + q
G_{ij}(\Phi,t)A^i(\Phi,t)  D_t\Phi^j- q A_0(\Phi,t)  \nonumber\\ 
&-&\!\!\!\!qG_{0i}(\Phi,t) D_t\Phi^i A_0(\Phi,t)
+mG(\Phi,t)G(\Phi,t)[\Phi,\Phi]^2+
(1-G_{00}(\Phi,t))+\cdots\bigg),
\eea 
in which $D_t=\partial_t+i[a_t(t),\;\;]$ acts as covariant derivative on
the world-line, and we have set the charge $\mu_0=q$. In above, the
functional dependence on the matrix coordinates of D0-branes should be
understood.  After all above, we have the action (\ref{SD0-general}) 
which can be interpreted as the world-line formulation of the dynamics of
D0-branes in nontrivial backgrounds.

\subsection{Action Of D0-Branes In One-Form Background}
In the following considerations in this work, we take the special case of
the dynamics of D0-branes in the background of 1-form RR field $(A_0(x,t),
A_i(x,t))$, in flat metric and zero NS B-field. Consequently, the low
energy bosonic action of $N$ D0-branes, after restoring
the string length $l$, is given by
\bea\label{SD0}
S_{{\rm D0}}= \int dt\; \tr \bigg(\haf m D_tX_i
D_t X^i + q D_tX^i A_i(X,t) - q A_0(X,t)+m
\frac{[X^i,X^j]^2}{4 (2\pi l^2)^2}+
\cdots\bigg),
\eea
in which we have changed slightly the notation for matrix coordinates
from $(2\pi l^2)\Phi^i$ to $X^i$, with the usual expansion
\bea\label{expan}
X^i=X^i_a T^a, \;\;\;i=1,\cdots, d,\;\;\;a=0,1,\cdots, N^2-1,
\eea
with $T^a$ as the basis for hermitian matrices (\ie the generators of
$U(N)$). Though D$p$-branes of String Theory live in the critical
dimensions $D=10$ (or 26), for the case of D0-branes it will be useful to
consider the more general case in arbitrary spatial dimensions $d$. We
recall that the gauge fields appear in the action by functional dependence
on symmetrized products of matrix coordinates $X$'s. The action
(\ref{SD0}) can be interpreted as the world-line formulation of
``electrodynamics on matrix space"  \cite{0104210}. Also we mention that
in this action the degrees of freedom is enhanced from $d$ in ordinary
space, to $d\times N^2$ in space with matrix coordinates. 

The original theory, which may be called bulk theory, is invariant under
the usual $U(1)$ transformations such as
\bea
A_\mu(x,t) \goes
A'_\mu(x,t)=A_\mu(x,t)-\partial_\mu\Lambda(x,t),\;\;\;\mu=0,1,\cdots,d
\eea
In the world-line theory, the transformation takes the form as: 
\bea\label{UOT}
A_i(X,t)&\goes& A'_i(X,t)=A_i(X,t)+\delta_i\Lambda(X,t),\nonumber\\
A_0(X,t)&\goes& A'_0(X,t)=A_0(X,t)-\partial_t\Lambda(X,t)
\eea
in which $\delta_i$ is the functional derivative $\frac{\delta}{\delta
X^i}$. Consequently, one obtains:
\bea\label{VSD0}
\delta S_{{\rm D0}} 
&\sim& q \int dt\; \tr \bigg(
\partial_t \Lambda(X,t) +
\dot X^i \delta_i\Lambda(X,t) +
i a_t[X^i,\delta_i\Lambda(X,t)]\bigg)\nonumber\\
&\sim& q \int dt\; \tr \bigg(
\frac{d\Lambda(X,t)}{dt}  +
i a_t[X^i,\delta_i\Lambda(X,t)]\bigg)\sim 0.
\eea
In above, the first term gives a surface term, and the second term
vanishes by the symmetrization prescription \cite{0103262} \footnote{The
general proof of invariance of the full Chern-Simons action is reported in
\cite{0105253} recently.}.

The equations of motion for $X$'s and $a_t$ by action (\ref{SD0}), after
ignoring for the moment the commutator potential $[X_i,X_j]^2$, will be
found to be \cite{0103262,0104210}
\bea\label{LORENZ}
&~&mD_tD_t X_i=q\bigg(E_i(X,t)
+\underbrace{D_tX^jB_{ji}(X,t)}\bigg),\\
\label{A0EOM}
&~&m[X_i,D_tX^i]=q[A_i(X,t),X^i],
\eea
with the following definitions
\bea
\label{ELEC}
E_i(X,t)&\equiv&-\delta_i A_0(X,t)-\partial_t A_i(X,t),\\
\label{MAGN}
B_{ji}(X,t)&\equiv&-\delta_jA_i(X,t)+\delta_iA_j(X,t).
\eea
In (\ref{LORENZ}), the symbol $\underbrace{D_tX^j B_{ji}(X,t)}$ denotes
the average over all of positions of $D_tX^j$ between the $X$'s of
$B_{ji}(X,t)$. The above equations for the $X$'s are like the Lorentz
equations of motion, with the exceptions that two sides are $N\times N$
matrices, and the time derivative $\partial_t$ is replaced by its
covariant counterpart $D_t$ \cite{hulldorn}. 

An equation of motion, similar to (\ref{LORENZ}), is considered in
\cite{fat241,fat021} as a part of similarities between the dynamics of
D0-branes and bound states of quarks--QCD strings in a baryonic state
\cite{fat241,fat021,02414}. The point is that, the dynamics of the bound
state center-of-mass (c.m.) is not affected directly by the non-Abelian
sector of the background, \ie the c.m. is ``white" with respect to $SU(N)$
sector of $U(N)$. The c.m. coordinates and momenta are defined by: 
\bea\label{CM}
X^i_\cm\equiv \frac{1}{N}
\tr X^i,\;\;\;\;  P^i_\cm\equiv \tr P^i,
\eea
where we are using the convention $\tr {\bf 1}_N=N$. To specify the net
charge of a bound state, as an extended object, its dynamics should be
studied in zero magnetic and uniform electric fields, \ie$B_{ji}=0$ and
$E_i(X,t)=E_{0i}$ \footnote{In a non-Abelian gauge theory a uniform
electric field can be defined up to a gauge transformation, which is quite
well for identification of white (singlet) states.}. Since the fields are
uniform, they are not involved by $X$ matrices, and contain just the
$U(1)$ part. In other words, under gauge transformations $E_{0i}$ and
$B_{ji}=0$ transform to $\tilde{E}_i(X,t)=V^\dagger(X,t)E_{0i}V(X,t)
=E_{0i}$ and $\tilde{B}_{ji}= 0$. Thus the action (\ref{SD0}) yields the
following equation of motion: 
\bea
(Nm){\ddot{X}}^i_\cm=Nq E^i_{0(1)},
\eea
in which the subscript (1) emphasises the $U(1)$ electric field. So the
c.m. interacts directly only with the $U(1)$ part of $U(N)$. From the
String Theory point of view, this observation is based on the simple fact
that the $SU(N)$ structure of D0-branes arises just from the internal
degrees of freedom inside the bound state. 

The world-line formulation we have here, is very much similar to the 
M(atrix) model conjecture; in particular, it is in the non-relativistic
limit. For the case of the dynamics of a charged particle with ordinary
coordinates, we can see easily that the light-cone dynamics have a form
similar to that we have in action (\ref{SD0}); see Appendix of
\cite{fat021}. To approach the Lorentz covariant formulation, following
finite-$N$ interpretation of \cite{9704080}, it is reasonable to interpret
things here in the DLCQ framework. This also should be applied for the
correspondence we consider to an effective field theory for the
interacting theory of D0-branes--photons. 

\subsection{Symmetry Transformations} 
Actually, the action (\ref{SD0}) is invariant under the transformations
\bea\label{Xat}
X^i&\goes&\tilde X^i=U^\dagger X^i U,\nonumber\\
a_t(t)&\goes& \tilde a_t(X,t)=U^\dagger a_t(t) U -i
U^\dagger \partial_t U,
\eea
with $U\equiv U(X,t)$ as an arbitrary $N\times N$ unitary matrix;
in fact under these transformations one obtains
\bea\label{DTF1}
D_tX^i&\goes&\tilde
D_t\tilde X^i=U^\dagger D_tX^iU,\\
\label{DTF2}
D_tD_tX^i&\goes&\tilde D_t \tilde D_t
\tilde X^i=U^\dagger D_tD_tX^iU.
\eea
Now, in the same spirit as for the previously introduced $U(1)$ symmetry of
eq.(\ref{UOT}), one finds the symmetry transformations: 
\bea\label{NAT}
X^i&\goes&\tilde X^i=U^\dagger X^i U,\nonumber\\
a_t(t)&\goes&\tilde a_t(X,t)=U^\dagger a_t(t) U -i
U^\dagger \partial_t U,\nonumber\\
A_i(X,t)&\goes& \tilde A_i(X,t)=
U^\dagger A_i(X,t)U+iU^\dagger\delta_i U,\nonumber\\
A_0(X,t)&\goes& \tilde A_0(X,t)=
U^\dagger A_0(X,t)U-iU^\dagger\partial_t U,
\eea
in which we assume that $U\equiv U(X,t)=\exp(-i\Lambda)$ is arbitrary up
to this condition that $\Lambda(X,t)$ is totally symmetrized in the $X$'s.
The above transformations on the gauge potentials are similar to those of
non-Abelian gauge theories, and we mention that it is just the consequence
of enhancement of degrees of freedom from numbers ($x$) to matrices ($X$). 
In other words, we are faced with a situation in which ``the rotation of
fields" is generated by ``the rotation of coordinates". The above
observation on gauge symmetry associated to D0-brane matrix coordinates,
on its own is not a new one, and we already know another example of this
kind in non-commutative gauge theories; see \cite{0104210}. In addition,
the case we see here for D0-branes may be considered as finite-$N$ version
of the relation between gauge symmetry transformations and transformations
of matrix coordinates \cite{0007023}.

The behaviour of eqs. (\ref{LORENZ}) and (\ref{A0EOM}) under gauge
transformation (\ref{NAT}) can be checked. Since the action is invariant
under (\ref{NAT}), it is expected that the equations of motion change
covariantly. The left-hand side of (\ref{LORENZ}) changes to $U^\dagger
D_tD_tX U$ by (\ref{DTF2}), and therefore we should find the same change
for the right-hand side. This is in fact the case, since for any function
$f(X,t)$ under transformations (\ref{Xat}) we have:
\bea
f(X,t) &\goes& \tilde{f}(\tilde{X},t)=
U^\dagger f(X,t)U,\nonumber\\
\delta_i f(X,t)&\goes& 
\tilde{\delta}_i\tilde{f}(\tilde{X},t)=
U^\dagger \delta_i f(X,t)U,\nonumber\\
\partial_t f(X,t)&\goes&
\partial_t \tilde{f}(\tilde{X},t)=
U^\dagger \partial_t f(X,t) U.
\eea
In conclusion, the definitions (\ref{ELEC}) and
(\ref{MAGN}), lead to
\bea\label{gtfs}
E_i(X,t) &\goes& \tilde{E_i}(\tilde{X},t)=
U^\dagger E_i(X,t)U,\nonumber\\
B_{ji}(X,t) &\goes& \tilde{B}_{ji}(\tilde{X},t)=
U^\dagger B_{ji}(X,t)U,
\eea
a result consistent with the fact that $E_i$ and $B_{ji}$ are functionals
of $X$'s. We thus see that, in spite of the absence of the usual
commutator term $i[A_\mu,A_\nu]$ of non-Abelian gauge theories, in our
case the field strengths transform like non-Abelian ones. We recall that
these all are consequences of the matrix coordinates of D0-branes. 
Finally by the similar reason for vanishing the second term of
(\ref{VSD0}), both sides of (\ref{A0EOM}) transform identically.

The last notable points are about the behaviour of $a_t(t)$ and $A_0(X,t)$
under symmetry transformations (\ref{NAT}). From the world-line theory
point of view, $a_t(t)$ is a dynamical variable, but $A_0(X,t)$ should be
treated as a part of background, however they behave similarly under
transformations. Also we see by (\ref{NAT}) that the coordinate
independence of $a_t(t)$, which is the consequence of dimensional
reduction, should be understood up to a gauge transformation. In
\cite{0103262} a possible map between the dynamics of D0-branes, and the
semi-classical dynamics of charged particles in Yang-Mills background was
mentioned. It is worth mentioning that via this possible relation, an
explanation for the above notable points can be recognized \cite{0103262}.

\section{Quantum Theory In One-Form Background}
\setcounter{equation}{0}
\subsection{Some General Aspects Of Bound State--Photon Interaction}
Before presenting the formulation, it is useful to mention some general
aspects of the problem at hand. At first let us recall another
representation of the symmetrization on the matrix coordinates. The other
useful symmetric expansion is done by using the Fourier components of a
function. To gain this Fourier expansion in matrix coordinates (we may
call it `non-Abelian Fourier expansion'), one can simply interpret the
derivatives of usual coordinates $\partial_{x^i}$ in (\ref{NTE}) as
momentum numbers $ik_i$. It is then not hard to see that for an arbitrary
function $f(X,t)$ the non-Abelian Fourier expansion will be found to be: 
\bea\label{NFE}
f(X,t)= \int d^dk\; \bar f (k,t)\;\e^{ik_i X^i},
\eea
in which $\bar f(k,t)$ are the Fourier components of the function
$f(x,t)$ (\ie function by ordinary coordinates) which is defined by the
known expression: 
\bea
\bar f (k,t) \equiv \frac{1}{(2\pi)^d}\int d^dx\; f(x,t)\; \e^{-ik_i x^i}.
\eea 
Since the momentum numbers $k_i$'s are ordinary numbers, and so commute
with each other, the symmetrization prescription is automatically
recovered in the expansion of the momentum eigen-functions $\e^{ik_i
X^i}$. This picture of symmetrization for matrix coordinates is similar to
that we already know for the Weyl ordering in phase space $(\hat q,\hat
p)$, with $[\hat q, \hat p]=i$.

Now, by using the symmetric expansion (\ref{NFE}), we can imagine some
general aspects of the interaction between D0-brane bound states and RR
photons. We recall that the bound state of D0-branes is described by the
action (\ref{SD0}) after setting $A_\mu(x,t)  \equiv 0$.  We mention that
still the degrees of freedom interact due to the commutator potential. By
doing a simple dimensional analysis it can be shown that the size scale of
the bound state, for finite number $N$ of D0-branes is finite and is of
order of $\ell\sim m^{-1/3} l^{2/3}$ \cite{9603081,fat021}. We recall that
the action we are using is coming from the string perturbative
calculations, and consequently we have for the size scale the further
relation $\ell \ll l$ \cite{9603081,fat021}.

Before proceeding further, we should distinguish the dynamics of the c.m. 
from the internal degrees of freedom of the bound state.  As mentioned in
before, the c.m. position and momentum of the bound state are presented by
the $U(1)$ sector of the $U(N)=SU(N)\times U(1)$, and thus the information
related to the c.m. can be gained simply by the $\tr$operation, relation
(\ref{CM}). So, the internal degrees of freedom of the bound state, which
consist the relative positions of $N$ D0-branes together with the dynamics
of strings stretched between D0-branes, are described by the $SU(N)$
sector of the matrix coordinates. It is easy to see that the commutator
potential in the action has some flat directions, along which the
eigen-values can take arbitrary large values. But it is understood that,
by considering the quantum effects and in the case that we expect
formation of the bound state, we should expect suppression the large
values of the internal degrees of freedom \cite{dewit-nicolai}.
Consequently, it is expected that the $SU(N)$ sector of matrix coordinates
take mean values like $\langle X^i_a \rangle \sim \ell$ ($a=1,\cdots,
N^2-1$, not $a=0$ as c.m.), with $\ell$ as the bound state size scale
mentioned in above \footnote{There is another way to justify this
expectation. It is known that diagonal $SU(N)$ matrices present the
relative positions of D0-branes, which are expected to be of order of
$\ell$ in a bound state. But due to the symmetry transformation we
introduced in previous section, the diagonal and non-diagonal elements in
the matrices can mix with each other, representing the same mechanical
system.  So, the size scale associated to the diagonal elements should be
valid also for the non-diagonal elements.}. We should mention that, though
the c.m. is represented by the $U(1)$ sector, but its dynamics is affected
by the interaction of the ingredients of bound state with the $SU(N)$
sector of external fields, similar to the situation we imagine in the case
of the Van der Waals force. 

The important question about the interaction of a bound state (as an
extended object) with an external field, is about `the regime in which the
substructure of bound state is probed'. As we mentioned in introduction,
in our case the quanta of RR fields are the representatives of the
external field. The quanta are coming from a `source' and so, as it makes
easier things, we ignore its dynamics. The source is introduced to our
problem by the gauge field $A_\mu(x,t)$. These fields appear in the action
by functional dependence on matrix coordinates $X$'s.  In fact this is the
key of how we can probe the substructure of the bound state. According to
the non-Abelian Fourier expansion we mentioned in above, we have
\bea 
A_\mu(X,t)= \int d^dk\; \bar A_\mu (k,t)\e^{ik_i X^i},
\eea 
in which $\bar A_\mu(k,t)$ is the Fourier components of the fields
$A_\mu(x,t)$ (\ie fields by ordinary coordinates). One can imagine the
scattering processes which are designed to probe inside the bound state.
Such as every other scattering process two limits of momentum modes,
corresponding to long and short wave-lengths, behave differently. 

\begin{figure}[t]
\begin{center}
\leavevmode
\epsfxsize=100mm
\epsfysize=60mm
\epsfbox{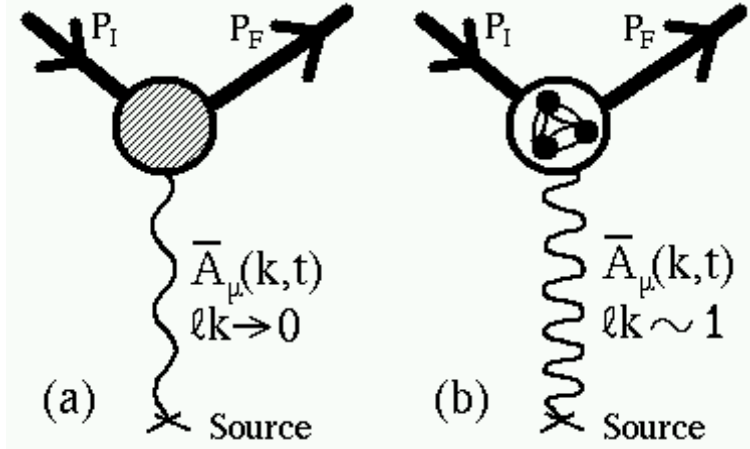}
\caption{Substructure is not seen by the long wave-length modes (a). Due
to functional dependence on matrix coordinates, the short wave-length
modes can probe inside the bound state (b). $\ell$ and $\bar A_\mu (k,t)$
represent the size of the bound state and the Fourier modes,
respectively.}
\end{center}
\end{figure}

In the limit $\ell |k|\goes 0$ (long wave-length regime), the field
$A_\mu$ is not involved by $X$ matrices mainly. It means that the fields
appear to be nearly constant inside the bound state, and in rough
estimation we have
\bea
\e^{ik_i X^i} \sim \e^{ik_i X^i_\cm}.
\eea
So in this limit we expect that the substructure and consequently
non-commutativity will not be seen; Figure-1a. As the consequence, after
interaction with a long wave-length mode, it is not expected that the
bound state jump to another energy level different from the first one. It
should be noted that the c.m. dynamics can be affected as well in this
case. 

In the limit $\ell |k|=$finite (short wave-length regime), the fields
depend on coordinates $X$ inside the bound state, and so the substructure
responsible for non-commutativity should be probed; Figure-1b. In fact, we
know that the non-commutativity of D0-brane coordinates is the consequence
of the strings which are stretched between D0-branes. So, by these kinds
of scattering processes, one should be able to probe both D0-branes (as
point-like objects), and the strings stretched between them. In this case,
it is completely expectable that the energy level of the incoming and
outgoing bound states will be different, since the ingredients of bound
state substructure can absorb quanta of energy from the incident wave. In
this case the c.m. dynamics can be affected in a novel way by the
interaction of the substructure with the external fields (the Van der
Waals effect). 

In general case, one can gain more information about the substructure of a
bound state by analysing the `recoil' effect on the source. To do this,
one should be able to include the dynamics of the source in the
formulation. Considering the dynamics of source, in the terms of quantized
field theory, means that we consider the processes in which the source and
the target exchange `one quanta of gauge field' with definite wave-length
and frequency, though off-shell, as $A_\mu(x,t) \sim \epsilon_\mu
\e^{ik_ix^i-i\omega t}$.  This kind of process is shown in Figure-2.

\begin{figure}[t]
\begin{center}
\leavevmode
\epsfxsize=45mm
\epsfysize=55mm
\epsfbox{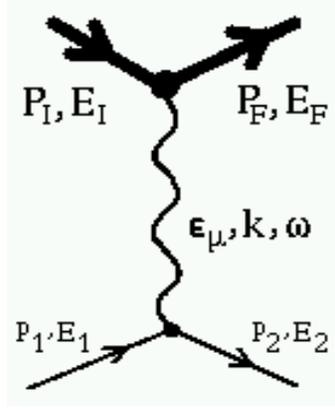}
\caption{Exchange of one photon between a D0-brane bound state (thick
lines) and another source (thin lines).}
\end{center}
\end{figure}

\subsection{Path Integral Quantization}
In this subsection we consider the quantisation of D0-brane dynamics,
using the path integral method. The theory on the world-line enjoys a
gauge symmetry, defined by the transformations (\ref{NAT}). We should fix
this symmetry, and here we use simply the temporal gauge, defined by the
condition $a_t(t)\equiv 0$. So after the Wick rotation $t\goes -it$ and
$A_0\goes -iA_0$, we have the following expression for the path integral
of our system:
\bea\label{path-int1}
\langle X_F,t_F | X_I, t_I \rangle \sim
\int [DX][Da_t]\; \delta(a_t)\; \det|\frac{\delta a_t}{\delta
\Lambda}|
\;\e^{-S_{{\rm D0}}[X,a_t]},
\eea
in which $\delta(a_t)$ supports the gauge fixing condition, and
$\det|\frac{\delta a_t}{\delta \Lambda}|$ is the determinant which arises
by variation of gauge fixing condition, and finally $S_{{\rm D0}}[X,a_t]$
is the action (\ref{SD0}) evaluated between $(X_F,t_F)$ and $(X_I,t_I)$,
as `Final' and `Initial' conditions.  The variation of gauge fixing
condition can be calculated in our case easily, and it is found to be (for
$U(X,t)=\exp(-i\Lambda)$): 
\bea
a_t=0\goes a'_t=\delta a_t=-i U^\dagger\partial_t U= -
\partial_t\Lambda(X,t)+ O(\Lambda^2),
\eea
and consequently we have $\frac{\delta a_t(t)}{\delta \Lambda(t')}
=-\partial_t\delta(t-t')$. So, we see that the determinant and
consequently the corresponding ghosts are decoupled from our dynamical
fields $X$'s \footnote{This case is similar to the so-called axial gauge
in extreme limit $\lambda\goes\infty$; page 196 of \cite{sterman}.}. So,
up to a normalization factor, we have for the above expression of path
integral:
\bea\label{path-int2}
\langle X_F,t_F | X_I, t_I \rangle \sim
\int [DX] \;\e^{-S_{{\rm D0}}[X,a_t\equiv 0]}.
\eea
To calculate the path integral in a general background we have to use the
perturbation expansion in the powers of charge $q$; this expansion is also
valid for the weak external fields $(A_0,A_i)$. So we have
\bea\label{path-int3}
\langle X_F,t_F | X_I, t_I \rangle &\sim& \int [DX]
\;\e^{-\int_{t_I}^{t_F} dt\; \tr (\haf m \dot{X}_i
\dot{X}^i +m\frac{[X^i,X^j]^2}{4 (2\pi l^2)^2})}
\nonumber\\
&~& \times \sum_{n=0}^\infty \frac{q^n}{n!}\;
\bigg\{i\int_{t_I}^{t_F} dt\; \tr\bigg(\dot{X}^i A_i(X,t) +
A_0(X,t)\bigg)\bigg\}^n.
\eea
As mentioned before, from the point of view of D0-brane dynamics,
the commutator potential $[X^i,X^j]^2$ is responsible for the formation of
D0-brane bound states \cite{9603081}. Though the problem of
finding the full set of eigen-energies and eigen-vectors of the
corresponding Hamiltonian is a very hard task, we assume that this full
set is at hand. It is logical to separate the c.m.  variables from the
internal ones; we show those of c.m. by momenta $P_\cm$ and
$|P_\cm\rangle$, and those of internal ones by energy $E_{\{n\}}$ and
$|\{n\}\rangle$, in which $\{n\}$ presents all of the quantum numbers
associated to the internal dynamics. We recall that the c.m. is free in
the case of $q=0$. It is worth to recall that, in general we expect that
the eigen-energies have the general form of $E_{\{n\}}=g(\{n\})
\ell^{-1}$, with $g(\{n\})$ as a function of quantum numbers $\{n\}$, and
as well the condition
\bea
\langle X | \{n\} \rangle \goes 0, \;\;\;\;\;\; {\rm for}\;\; |X|\gg \ell,
\eea
for the wave-functions, with $\ell\sim m^{-1/3} l^{2/3}$ the size scale of
the bound state we mentioned in before. As any other quantum mechanical
system, for the case $q=0$ the general expression of the propagator can be
used:
\bea\label{propagator}
\langle X_2,t_2 | X_1, t_1 \rangle_{q=0} = \sum_{P_\cm} 
\sum_{\{n\}} 
\langle X_2|P_\cm, \{n\}\rangle \langle P_\cm, \{n\}|X_1\rangle
\e^{-(\frac{P_\cm^2}{2Nm}+E_{\{n\}})(t_2-t_1)},
\eea
with definition $|P_\cm, \{n\}\rangle\equiv |P_\cm\rangle\otimes
|\{n\}\rangle$. We now can insert the propagator above in last expression
(\ref{path-int3}), with this care that the perturbation expansion have
terms involved by velocity $\dot X$. Based on the standard representation
of `slicing' we use for the path integrals, finally the following
expression for the first order of perturbation will be found (see
\cite{ryder}): 
\bea\label{path-int4}
\langle X_F,t_F | X_I, t_I \rangle &\sim& 
\langle X_F,t_F | X_I, t_I \rangle_{q=0} \nonumber\\
&+&i {\cal N}\lim_{\Delta t \goes 0} \sum_{k=1}^n \int
d^dX_{k-1} d^dX_k d^dX_{k+1}\times
\langle X_F,t_F | X_{k+1}, t_{k+1} \rangle_{q=0}\times\nonumber\\
&~&2\Delta t \cdot\tr\bigg(q \frac{X^i_{k+1} - X^i_{k-1}}{2\Delta t}
A_i(X_k,t) +qA_0(X_k,t)\bigg)\times \nonumber\\
&~&\langle X_{k-1},t_{k-1} | X_I, t_I \rangle_{q=0}\times \e^{-S_{q=0} [k,k-1;\Delta t]}\;\;
\e^{-S_{q=0} [k+1,k;\Delta t]} 
\nonumber\\&+& O(q^2),
\eea
in which $t_j-t_I=j\cdot\Delta t$ and $t_F-t_I=(n+1)\Delta t$. In above
$S_{q=0}[j,j+1;\Delta t]$ is the value of the action in the exponential of
(\ref{path-int3}) evaluated between the points $(X_j,t_j)$ and
$(X_{j+1},t_{j+1})$ ($1\leq j \leq n$) by limit $\Delta t\goes 0$. The
normalization constant ${\cal{N}}$ contains sufficient powers of $\Delta
t$ to make the final result finite and independent of $\Delta t$. The sum
$\sum_k$ is coming from slicing the potential term $\int dt\tr (\dot
X\cdot A + A_0)$ in path integral (\ref{path-int3}), and it eventually
will change to the time integral $\int dt$ over the intermediate `times'
in which the interaction occur. It is worth to recall that the spatial
integrals like $\int d^d X$ are in fact as $\int \prod_{a=0}^{N^2-1} d^d
X_a$. We mention that for the velocity independent term $A_0(X,t)$, the
integrals of $d^dX_{k\pm 1}$ can be done to get the new propagators, and
after the change $X_k\goes X$, we simply find the expression like
\bea
\sim i \int_{t_I}^{t_F} dt \int d^dX \;
\langle X_F,t_F | X, t \rangle_{q=0}
\tr\bigg(qA_0(X,t)\bigg) 
\langle X,t | X_I, t_I \rangle_{q=0}
+ O(q^2),
\eea
which is the familiar expression for the velocity independent
interactions. 

For many practical aims, we should find the S-matrix elements between
states with definite momenta and energies; Figure-3. This can be done by
the proper transformations on the amplitudes $\langle X_F,t_F | X_I, t_I
\rangle$ in the coordinate space. 

\begin{figure}[t]
\begin{center}
\leavevmode
\epsfxsize=50mm
\epsfysize=55mm
\epsfbox{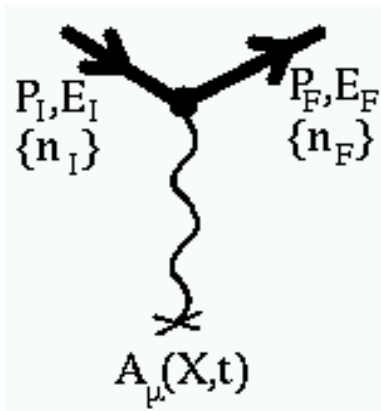}
\caption{The graph for the transition amplitude between states with
definite c.m. momenta and energies $(P,E)$, and internal energy specified
by the quantum numbers $\{n\}$.}
\end{center}
\end{figure}

Due to less knowledge about the propagator (\ref{propagator}), the last
expression (\ref{path-int4}) still can not be used for the actual
calculations. As mentioned in before, we expect that the spatial
integrations $\int d^d X$ find their main contribution from the volume of
bound state $V\sim \ell^d$. So as an approximation and to know a little
more about the result, we may ignore the commutator potential, but doing
integrations in the finite volume $V\sim \ell^d$, or in the case, we
simply put $\int d^dX_a\sim \ell^d$, for $a\neq 0 =$c.m. By doing this, we
can verify the general aspects about the probing of the substructure of
the bound state, discussed in the previous subsection.

\subsection{Effective Interaction Vertex Of Photon And `Free'\\ D0-Branes}
In the considerations of the previous subsection, the background
$(A_0(x,t), A_i(x,t))$ was taken to be arbitrary. Here we take the example
in which the D0-branes interact with a monotonic incident wave, defined by
the condition $\bar{A}_\mu (k',\omega') 
=\epsilon_\mu\delta^d(k'-k)\delta(\omega'- \omega) $, with $\epsilon_\mu$
as the polarization vector, and the following definition for the Fourier
modes: 
\bea
\bar{A}_\mu(k',\omega')  \equiv \frac{1}{(2\pi)^{d+1}}\int d^dx\; dt\; 
A_\mu(x,t)\; \e^{-ik_i' x^i+i\omega' t}.
\eea
So the corresponding gauge field is $A_{\mu}(X,t)\sim\epsilon_\mu
\exp(ik_i X^i-i\omega t)$. Besides, here we ignore the commutator
potential, and consequently it is assumed that all of the $N^2$ degrees of
freedom, including those $N$ ones which describe the position of
D0-branes, are free for $q=0$. So we have the following expression for the
path integral: 
\bea\label{path-int5}
\langle X_F,t_F | X_I, t_I \rangle &\sim& \int [DX]
\;\e^{-\int_{t_I}^{t_F} dt\; \tr (\haf m \dot{X}_i
\dot{X}^i)}
\nonumber\\
&~& \times 
\bigg\{1+iq \int_{t_I}^{t_F} dt\; \tr\bigg(\dot{X}^i A_i(X,t) +
A_0(X,t)\bigg)+O(q^2)\bigg\}.
\eea
A similar theory for a charged particle in ordinary space is considered in
the Appendix-A, to extract the field theory vertex function of the
coupling of a `photon' to incoming--outgoing charged particles. So the
result of the path integral above, can be considered as the `matrix
coordinate' version of the example of the Appendix-A. We continue with an
expression like that of (\ref{path-int4}), as:
\bea\label{path-int6}
\langle X_F,t_F | X_I, t_I \rangle &\sim& 
\langle X_F,t_F | X_I, t_I \rangle_\fp \nonumber\\
&+& i{\cal N}\lim_{\Delta t \goes 0} \sum_{k=1}^n \int
d^dX_{k-1} d^dX_k d^dX_{k+1}\times
\langle X_F,t_F | X_{k+1}, t_{k+1} \rangle_\fp\times\nonumber\\
&~&2\Delta t \cdot\tr\bigg(q \frac{X^i_{k+1} - X^i_{k-1}}{2\Delta t}
A_i(X_k,t) +qA_0(X_k,t)\bigg)\times \nonumber\\
&~&\langle X_{k-1},t_{k-1} | X_I, t_I \rangle_\fp\times \e^{-S_\fp
[k,k-1;\Delta t]}\;\;
\e^{-S_\fp [k+1,k;\Delta t]} 
\nonumber\\&+& O(q^2),
\eea
in which $S_\fp$ and $\langle \cdots \rangle_\fp$ are the action and the
propagator of free particles, respectively; see Appendix-B for the
explicit expressions. The integrations $d^dX_{k\pm 1}$ can be done to get
new propagators, and after the change $X_k\goes X$, we find: 
\bea\label{path-int7}
\langle X_F,t_F | X_I, t_I \rangle &\sim& 
\langle X_F,t_F | X_I, t_I \rangle_\fp\nonumber\\
&+& i{\cal N}' \int_{t_I}^{t_F} dt \int d^dX
\;\;\langle X_F,t_F | X, t \rangle_\fp\times
\nonumber\\
&~&\tr\bigg\{q\bigg(\frac{X^i_F - X^i}{t_F-t}+ \frac{X^i-X^i_I}{t-t_I}\bigg)
A_i(X,t) +qA_0(X,t)\bigg\}\times \nonumber\\
&~&\langle X,t| X_I, t_I \rangle_\fp + O(q^2).
\eea

\begin{figure}[t]
\begin{center}
\leavevmode
\epsfxsize=50mm
\epsfysize=55mm
\epsfbox{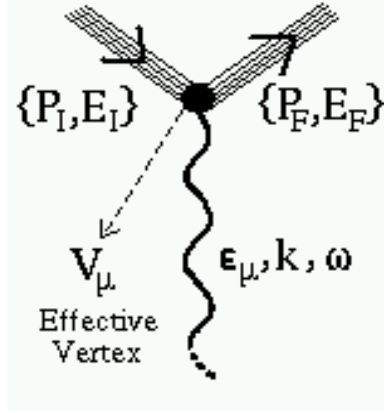}
\caption{The graph for the transition amplitude between states with
definite momenta and energies, specified with the set $\{P,E\}$, for all
of $N^2$ degrees of freedom. Here we use some thin lines as incoming and
outgoing states, to emphasize that these states are free before and after
the vertex of interaction.}
\end{center}
\end{figure}
Up to now the gauge field can be in any arbitrary form. Also, since in
this case we have ignored the commutator potential and so the degrees of
freedom are free for $q=0$, we can easily use the momentum basis for the
incoming-outgoing states; see Figure-4. So for the S-matrix element in
momentum-energy basis, we have the expression: 
\bea\label{amplitude-1}
S_{FI}&\sim& \prod_{a=0}^{N^2-1} \delta^d(P_{Fa}-P_{Ia})
\delta(E_{Fa}-E_{Ia})\nonumber\\
&+& (\cdots) \int_{t_I}^{t_F}dt\int d^d X 
\int d^dX_I\; d^dX_F
\prod_{a=0}^{N^2-1}\bigg( \e^{i(E_{Fa}t_F-E_{Ia}t_I)}
\e^{-i(P_{Fa}\cdot X_{Fa}-P_{Ia}\cdot X_{Ia})}\bigg)\nonumber\\
&~&\langle X_F,t_F | X, t \rangle_\fp\nonumber\\
&~&\tr\bigg(
iq \epsilon\cdot \bigg(\frac{X_F - X}{t_F-t}+
\frac{X-X_I}{t-t_I}\bigg)\e^{ik\cdot X-i\omega t}
+iq\epsilon_0 \e^{ik\cdot X-i\omega t}
\bigg)  \nonumber\\
&~&\langle X,t| X_I, t_I \rangle_\fp+O(q^2),
\eea
in which $E_a=P_a^2/(2Nm)$ for both $I$ and $F$ states (by convention
$\tr(T^aT^b)=N\delta^{ab}$), and the symbol $A\cdot B$ is for the inner
product $A_iB^i$. We recall that the subscripts $a$ and $b$ are counting
the $N^2$ independent degrees of freedom associated with $N\times N$
hermitian matrices $X$ or $P$. Some of the integrations above can be done
(see Appendix-B), and the resulting expression will be found to be: 
\bea\label{amplitude-2}
S_{FI}&\sim&\prod_{a=0}^{N^2-1}\delta^d(P_{Fa}-P_{Ia})
\delta(E_{Fa}-E_{Ia})\nonumber\\
&+&(\cdots)
\delta^d (P_{F\cm}-P_{I\cm}-k)
\delta\bigg(\sum_{a=0}^{N^2-1}(E_{Fa}-E_{Ia})-\omega\bigg)
\nonumber\\&\times&
\int \prod_{b=1}^{N^2-1} d^d\hat{X}_b \;
\e^{i(P_{Ib}-P_{Fb})\cdot \hat{X}_b}
\tr\bigg(
iq\bigg(\epsilon\cdot (P_F+P_I)+\epsilon_0\bigg) \e^{ik\cdot \hat{X}}
\bigg)+ O(q^2),
\eea
in which the second series of $\delta$-functions have appeared as supports
of the total momentum and total energy conservations. The last expression
contains $\tr$and integrals over the matrix coordinates $\hat{X}$
($\tr(\hat{X})=0$), and though the improved forms in some special cases
($N$=2 or in large-$N$ limit) are accessible, the result in general case
is not known. We mention that such integrals for ordinary coordinates, as
that of Appendix-A, can be calculated exactly. We can present the general
form of the result as:
\bea
S_{FI}&\sim&\prod_{a=0}^{N^2-1}\delta^d(P_{Fa}-P_{Ia})
\delta(E_{Fa}-E_{Ia})\nonumber\\
&+& (\cdots)
\delta^d (P_{F\cm}-P_{I\cm}-k)
\delta\bigg(\sum_{a=0}^{N^2-1}(E_{Fa}-E_{Ia})-\omega\bigg)
\nonumber\\
&~&\bigg(iq \epsilon\cdot
V(P_{Ia,Fa},k)+iq\epsilon_0V_0(P_{Ia,Fa},k)\bigg)
+O(q^2).
\eea
in which $V^\mu(P_{Ia,Fa},k)$, as the effective vertex function (see
Figure-4), has the general form:
\bea
&~&V^i=\tr\bigg((P^i_F+P^i_I) H(P_{Ia,Fa},k)\bigg)\nonumber\\
&~&V^0=\tr\bigg( H(P_{Ia,Fa},k)\bigg),
\eea
with $H(P_{Ia,Fa},k)$ as a matrix, depending on $P_{Ia}$, $P_{Fa}$
($a=1,\cdots, N^2-1$, $a\neq\cm$) and $k$. In the case of ordinary
coordinates for covariant theory we find simply $V^\mu\sim (p_I+p_F)^\mu$;
see Appendix-A. 

\section{Which Field Theory?}
\setcounter{equation}{0}
By the studies like those of \cite{9205205}, it is understood that the
world-line theory of a charged particle in ordinary space, in presence of
the gauge field background $A_\mu(x)$ can be corresponded to the second
quantized field theory of interaction of charges and photons, something
similar to the theory we imagine for the case of interaction of electrons
and photons. The corresponding action may be presented as (with $\dot
x=dx/d\tau$) 
\bea\label{ord-space}
S=\int d\tau\; (\haf m \dot x^2 -  q \dot x^\mu A_\mu(x)).
\eea
As an example, in the Appendix-A we derived the field theory vertex
function for the interaction of a photon with the currents of
incoming--outgoing charged particles. In the previous section we developed
the basic elements of the world-line formulation of the interaction of
D0-brane bound states with 1-form RR photons. Particularly, we showed how
various amplitudes can be calculated in principle by the world-line
theory, at least in the perturbative regime. In this section we want to
discuss `which field theory' can be corresponded to the amplitudes,
calculated by the world-line theory. This is like the same relation that
we consider in String Theory, between field theories in space-time and
theories which are living on the world-sheet of strings. As we saw in
previous section, our knowledge about the exact values of the amplitudes
is restricted, and hence the discussion here will be based on some
qualitative considerations. It remains for future studies to check the
relation quantitavely, in particular by comparing the amplitudes as
observable quantities. 

Probably one of the most guiding observations is the `matrix' nature of
the gauge fields in the world-line formulation. Due to functional
dependence of the gauge fields on matrix coordinates, the gauge fields
$A^\mu(X)$ in our theory are $N\times N$ hermitian matrices, and so the
gauge fields have the usual expansion in the matrix basis:
\bea\label{comp}
A^\mu(X)=A^\mu_a(X_b)T^a,\;\;\;\;\;\;\; 
A^\mu_a(X_b)\equiv\frac{1}{N} \tr (A^\mu(X) T^a),
\eea
in which $A^\mu_a(X_b)$ are some functions (numbers) depending on the
components of the matrix coordinates. The most famous `matrix' gauge
fields we know are those of non-Abelian gauge theories. Now, it is
tempting to see what kind of relation between these two kinds of matrix
gauge fields can be verified.

The best base we found for the possible relation mentioned in above is the
suggested relation of \cite{9908142}, as the map between field
configurations of non-commutative and ordinary gauge theories. The
suggested map preserves the gauge equivalence relation, and it is
emphasised that, due to different natures of the gauge groups, this map
can not be an isomorphism between the gauge groups. Now in our case, it
will be interesting to study the properties of the map between non-Abelian
gauge theory and gauge theory associated with matrix coordinates; on one
side the quantum theory of matrix fields, and on the other side the
quantum mechanics of matrix coordinates. It will be helpful to do some
imaginations in this direction. Since for the consideration we have in
below there is no essential difference between fermions and bosons, we
take the example of the interaction of a fermionic matter field with the
external non-Abelian gauge field $\ca_\mu(x)$, which is described by the
action
\bea
&~&S=\int d^{d+1}x\; \bigg(\bar\psi (i\gamma^\mu 
\partial_\mu -m)\psi-g\tr(J^\mu \ca_\mu)\bigg),\\
&~&\ca^\mu(x)=\ca^\mu_a(x)T^a,\nonumber
\eea
in which the term $J^\mu \ca_\mu$ is responsible for the interaction; it
may be chosen as that of the minimal coupling $J_\mu^a=i\bar\psi\gamma_\mu
T^a \psi$. Gauge invariance specifies the behaviour of the current $J_\mu$
under the gauge transformations to be $J(x)\goes J'(x)=U^\dagger J(x) U$. 
Though here we are treating the gauge field as a fixed background, in
general we can add the kinetic term of gauge fields by the action: 
\bea
&~&S_{{\rm gauge}}=\int d^{d+1}x\;
\tr(\frac{-1}{4}\cf^{\mu\nu}\cf_{\mu\nu}),\\ 
&~&\cf^{\mu\nu}(x)=\cf^{\mu\nu}_a(x)T^a,\;\;
\cf_{\mu\nu}=[\cd_\mu,\cd_\nu],\nonumber
\eea
with definition $\cd_\mu\equiv \partial_\mu-ig\ca_\mu$.  On the world-line
theory side, we have the theory of matrix coordinates. Let us, for the
moment, forget that the world-line theory considered so far is in the
non-relativistic limit, and consider things in a covariant theory. So, one
may have something like the action below in the world-line theory: 
\bea\label{mat-space}
S[X]=\int d\tau\;\tr \bigg(\haf m D_\tau X_\mu D_\tau X^\mu - q D_\tau
X^\mu A_\mu(X)+\cdots\bigg),
\eea
in which to make things easier, we have dropped any kind of potentials,
including the commutator potential of D0-branes. In above, $\tau$ is
parametrizing the world-line, $D_\tau=\partial_\tau+i[a_\tau,\;]$ is the
covariant derivative along the world-line, and $a_\tau$ is the world-line
gauge field \footnote{See \cite{0103262} for an example of these objects
in a covariant theory.}. The gauge field $A(X)$ depends on the symmetrized
products of $X$'s. In the same spirit of the transformations in the
world-line theory of D0-branes, we can take
\bea
X^\mu&\goes&\tilde X^\mu=U^\dagger X^\mu U,\nonumber\\
a_\tau&\goes&\tilde a_\tau=U^\dagger a_\tau U -i
U^\dagger \partial_\tau U,\nonumber\\
A_\mu(X)&\goes& \tilde A_\mu(\tilde X)=
U^\dagger A_\mu(X)U-iU^\dagger
\delta_\mu U,
\eea
as the gauge transformation in the covariant theory, with
$U=\exp(-i\Lambda(X,\tau))$. We mention that, $D_\tau X^\mu$ transforms as
follows under the transformation: $D_\tau X^\mu \goes\tilde D_\tau\tilde
X^\mu=U^\dagger D_\tau X^\mu U$. Following relations (\ref{ELEC}) and
(\ref{MAGN}), we can define the field strength as follows: 
\bea
F_{\mu\nu}(X)\equiv \delta_\mu A_\nu(X) - \delta_\nu A_\mu(X),
\eea
and so the field strength transforms as: $F_{\mu\nu}\goes\tilde
F_{\mu\nu}=U^\dagger F_{\mu\nu} U$; see equation (\ref{gtfs}). Now, we
want to sketch the map between the field theory in space-time and the
world-line theory of a charged particle in a matrix space. It is natural
to assume that the map should relate the objects in two theories as is
shown in the Table-1. 
\begin{table}[t]
\begin{center}
\begin{tabular}{c c c}
Non-Abelian Field Theory & {\large $\Leftrightarrow$} & 
Gauge Theory On Matrix Space\\\hline
$\ca^\mu(x)=\ca^\mu_a(x)T^a$ & {\large $\sim$} & $A^\mu(X) + $
gauge trans. terms\\
$\cf^{\mu\nu}(x)=\cf^{\mu\nu}_a(x)T^a$ & {\large $\sim$} &
$F^{\mu\nu}(X)$\\
$J^\mu(x)=J^\mu_a(x) T^a$ &  {\large $\sim$} & $D_\tau X^\mu$\\
$\Lambda(x)=\Lambda_a(x)T^a$ &{\large $\sim$} & $\Lambda(X)$
\end{tabular}
\caption{The quantities which should be related by the map in two
theories. $\Lambda$ is the symbol for the gauge transformation
parameter in two theories.}
\end{center}
\end{table}
We mention, 1) it is enough that the gauge fields are related up to
a gauge transformation, 2) the objects in both sides are matrices, and 3)
the field strengths and currents of the two theories transform
identically under the gauge transformations. 

Since in this case, we have matrices of equal sizes in both sides, it may
be considered as a case in which one is able to find a one-to-one map
between two theories. In particular, the one-to-one correspondence between
the currents of two theories, \ie $J^\mu$ and $D_\tau X^\mu$, suggests to
verify whether such a one-to-one map can be defined or not. The case for
the gauge fields $\ca_\mu(x)$ and $A_\mu(X)$ appears to be a little more
subtle, because, though the gauge fields in both sides are matrices, but
the numbers of the independent functions of gauge fields do not match. In
the side of ordinary gauge fields $\ca_\mu(x)$, we have $N^2\cdot (d+1)$
independent functions. In the matrix coordinates side, we know how to
construct functionals on matrix space: once a function on ordinary space
as $f(x)$ is introduced, we can find its matrix coordinates version $f(X)$
by the non-Abelian Taylor or Fourier expansions, relations (\ref{NTE}) and
(\ref{NFE}). So in the matrix coordinates side, the gauge field $A_\mu(X)$
is constructed just by $1\cdot (d+1)$ functions. In other words, in the
matrix theory side, since the gauge fields $A_\mu(X)$ are matrices due to
their functional dependence on matrix coordinates $X$, all of the
components of the gauge field, defined in (\ref{comp}), are non-zero and
specified in each direction by one function. One nice example is the plane
wave of Subsec.3.3. The counting above will be modified by considering the
gauge symmetry, but the mismatch will not be removed. So by this way of
counting, finding a one-to-one map seems to be out of hand. It remains for
future studies, to see whether such a mismatch between the number of the
functions can find an explanation, to provide the definition of a
one-to-one map \footnote{One suggestion can be looking for a one-to-one
map in the case which the ordinary gauge theory is in the ordered phase. 
From our statistical mechanics experience in transition from disordered
phase, in which the system is described by the most number of data, to
ordered one, we know the situations in which the reduction of the degrees
of freedom is the result of the phase transition. Particularly, the phase
in which all of the components of gauge field $\ca(x)$ and current $J(x)$
appear equivalently, can be considered as a situation in which the
one-to-one map can be defined.}. 

In \cite{0104210} a conceptual relation between the above map and the
ideas concerned in special relativity is mentioned; see also
\cite{fat241,fat021,02414}. According to an interpretation of the special
relativity program, it can be meaningful if the `coordinates' and the
`fields' in our physical theories have some kinds of similar characters. 
As an example, we observe that both the space-time coordinates $x^\mu$ and
the electromagnetic potentials $A^\mu(x)$ transform equivalently (\ie as a
$(d+1)$-vector) under the boost transformations. Also, by this way of
interpretation, the super-space formulations of supersymmetric field and
superstring theories are the natural continuation of the special
relativity program. In the case of above mentioned map, it may be argued
that the relation between `matrix coordinates' and `matrix fields' (gauge
fields of a non-Abelian gauge theory) is one of the expectations which is
supported by the spirit of the special relativity program. We recall that
the symmetry transformations of gauge theory on matrix space appeared to
be similar to those of non-Abelian gauge theories, relations (\ref{NAT}) 
and (\ref{gtfs}).

Finally, we should recall that, the theory which was considered in
previous sections, was a non-relativistic theory. As noted before, to
approach a covariant theory, the natural assumption is to interpret things
as the light-cone gauge formulation of a covariant theory. It is the same
approach by M(atrix) model conjecture \cite{9610043}, and in particular
its finite-$N$ version \cite{9704080}. Also the map we discussed in above
between the D0-brane theory and gauge theory, should be considered for the
light-cone formulation of the gauge theory side \cite{light-conr-ref}. In
the light-cone gauge formulation the non-relativistic mass $m$ is
interpreted as the unit of longitudinal momentum, as $p_+=m$.  Also we
mention that, to match the amplitudes by world-line formulation with
amplitudes by the field theory, the normalizations of the wave functions
are different between the non-relativistic and light-cone gauge
interpretations. 

\section{Conclusion And Discussion} 
\setcounter{equation}{0} 
In this work we provide the basic elements of the interaction of D0-brane
bound states and 1-form RR photons, by the world-line formulation. At the
classical level, we checked that the action is invariant under the gauge
transformation of the gauge fields in the bulk theory. Also, due to matrix
nature of the coordinates, we see that a new symmetry transformations
exist, which under these new ones the gauge fields transform as gauge
fields of a non-Abelian gauge theory. We interpret this observation as the
case in which ``the fields rotate due to rotation of coordinates". We
derived the Lorentz-like equations of motion, and the covariance of the
equations are checked under the symmetry transformations. It is seen that
the c.m. is `white' or `colourless' with respect to the $SU(N)$ sector of
the background fields. 

At the quantum level, we developed the perturbation theory of the
interaction of D0-branes with the RR gauge fields. In particular, using
the path integral method, we write down the expression of the propagator
in the first order of perturbation, which can be converted to the
amplitudes of the scattering processes by an arbitrary external source. 
It is discussed that how the functional dependence of gauge fields
provides the base for probing the substructure of the bound states.

We discussed the possibility that the theory on the world-line of
D0-branes can be mapped, maybe by a one-to-one map, to the non-Abelian
gauge field theory.

One natural extension of the studies in this work can be for the
supersymmetric case. Particularly in the case of maximal supersymmetry
($d=9$), we have the D0-branes of M(atrix) model, coupled to the 1-form RR
background. As mentioned in Introduction, in the M(atrix) model picture
D0-branes are assumed to be the super-gravitons of the 11-dimensional
super-gravity in the light-cone gauge, and in particular they play in the
case the role of the `photons' of the 1-form RR field in 10 dimensions. 
The interaction of one D0-brane with a bound state of D0-branes is studied
in the context of M(atrix) model, and according to the M(atrix) model
interpretation, the commutator potential is responsible for the
interaction of the single D0-brane (maybe viewed as one RR photon) and the
bound state. The known results are those of different orders of loop
calculations. It will be interesting to check whether the perturbation
expansion in charge $q$ of this work, can reproduce the loop
expansion results of M(atrix) model.

Another extension of the studies of this work, can be for including the
gravitational effects, specificaly by considering non-flat metrics.  The
comparison to the M(atrix) model calculations also can be done in this
case. 

\vspace{.5cm}

{\bf Acknowledgement:} The author is grateful to Theory Group of INFN
Section in `Tor Vergata University', specially to A. Sagnotti, for kind
hospitality. The careful readings of manuscript by M. Hajirahimi, and
specially S. Parvizi are acknowledged. The author uses the grant under the
executive letter no. $\frac{3/3746}{79/9/7}$, by the Ministry of Science,
Research and Technology of Iran.

\appendix
\section{Perturbation Theory Of A Charged Particle In
\\Ordinary Space By Path Integral Method}
\setcounter{equation}{0}
As an exercise, and to complete the basics of present paper, here we
review the perturbation theory of a charged particle in electromagnetic
background. Particularly, we extract the vertex function of the coupling
of a photon to incoming--outgoing (bosonic) charged particles. In contrast
to the non-relativistic theory of the paper, here we consider the
covariant example. A good reference for these discussions is \cite{ryder}.
The action we use, initially in Euclidean space-time, is simply
\bea
S=\int d\tau \;(\haf m\dot x^2 - i q \dot x^\mu A_\mu(x)),
\eea
We begin with the expression similar to the formula
(\ref{path-int4}) of the text as:
\bea\label{path-app1}
\langle x_F,\tau_F | x_I, \tau_I \rangle &\sim& 
\langle x_F,\tau_F | x_I, \tau_I \rangle_\fp \nonumber\\
&+&i {\cal N}\lim_{\Delta\tau \goes 0} \sum_{k=1}^n \int
d^{d+1}x_{k-1} d^{d+1}x_k d^{d+1}x_{k+1}\times
\langle x_F,\tau_F | x_{k+1}, \tau_{k+1} \rangle_\fp\times\nonumber\\
&~&2\Delta \tau \cdot\bigg(q \frac{x^\mu_{k+1} - x^\mu_{k-1}}{2\Delta\tau}
A_\mu(x_k,\tau)\bigg)\times 
\langle x_{k-1},\tau_{k-1} | x_I, \tau_I \rangle_\fp\times \nonumber\\
&~&\e^{-S_\fp [k,k-1;\Delta\tau]}\;\;
\e^{-S_\fp [k+1,k;\Delta\tau]} + O(q^2),
\eea
in which the normalization constant ${\cal N}$ contains sufficient powers
of $\Delta\tau$ to regulate the final result, and we have the following
relations:
\bea
&~&\langle x_2,\tau_2 | x_1, \tau_1 \rangle_\fp\sim
\e^{\frac{-m(x_2-x_1)^2}{2(\tau_2-\tau_1)}}\sim
\int d^{d+1}l\;
\exp\bigg(il\cdot(x_2-x_1)-i\frac{l^2}{2m}(\tau_2-\tau_1)\bigg),
\\
&~&S_\fp [j+1,j;\Delta\tau]=\frac{m(x_{j+1}-x_j)^2}{2\Delta\tau},
\eea
with $A\cdot B= A^\mu B_\mu$. Doing integrations $dx_{k\pm 1}$ to replace
new propagators, and after the change $x_k\goes x$, we will find: 
\bea\label{path-app2}
\langle x_F,\tau_F | x_I, \tau_I \rangle &\sim& 
\langle x_F,\tau_F | x_I, \tau_I \rangle_\fp
+ i{\cal N}' \int_{\tau_I}^{\tau_F} d\tau \int d^{d+1}x
\;\langle x_F,\tau_F | x, \tau \rangle_\fp\times
\nonumber\\
&~&\bigg\{q\bigg(\frac{x_F - x}{\tau_F-\tau}+
\frac{x-x_I}{\tau-\tau_I}\bigg)\cdot
A(x)\bigg\}\times 
\langle x,\tau| x_I, \tau_I \rangle_\fp + O(q^2),
\eea
From now on we restrict the calculation to the plane wave $A_\mu(x)\sim
\epsilon_\mu \e^{ik_\nu x^\nu}$.  To find the S-matrix elements, it is
usual to go to the momentum space, and we have the expression 
\bea
S_{FI}&\sim& \delta^d(p_F-p_I) \delta(E_F-E_I)+ 
\frac{{\cal N}''\e^{-im^2(\tau_F-\tau_I)/2}}{\tau_F-\tau_I}
\int_{\tau_I}^{\tau_F}
d\tau\int d^{d+1} x \int d^{d+1}x_I \int d^{d+1}x_F\nonumber\\
&~&\e^{-ip_F\cdot x_F} \e^{ip_I\cdot x_I}
\langle x_F,\tau_F | x, \tau \rangle_\fp
\bigg\{
\e^{ik\cdot x} 
\exp\bigg(iq \epsilon\cdot \bigg(\frac{x_F - x}{\tau_F-\tau}+
\frac{x-x_I}{\tau-\tau_I}\bigg)\bigg)
\bigg\}_{{\rm linear\;in}\;\epsilon}
\nonumber\\
&~&\langle x,\tau| x_I, \tau_I \rangle_\fp+O(q^2),
\eea
in which $p_F^2=p_I^2=-m^2$, and to make easier the calculation we have
exponentiated the $\epsilon_\mu$;  so we should keep only the linear term
in $\epsilon$ finally.  By using the momentum representation of the
propagator $\langle \cdots
\rangle_\fp$ we find:
\bea
S_{FI}&\sim& \delta^d(p_F-p_I) \delta(E_F-E_I)\nonumber\\
&+& (\cdots)
\delta^d(p_F-p_I-k) \delta(E_F-E_I-k_0)(iq \epsilon_\mu (p_F+p_I)^\mu)
+O(q^2),
\eea
in which we recognize the field theory result $\epsilon\cdot (p_F+p_I)$
for the vertex function (page 548 of \cite{sterman}). 

\section{Calculation Of S-Matrix Element For Matrix
\\Coordinates In Momentum Basis}
\setcounter{equation}{0}
Here we present the derivation of (\ref{amplitude-2}), starting with
(\ref{amplitude-1}). By using the definitions: 
\bea
&~&\langle X_2,t_2 | X_1, t_1 \rangle_\fp\sim
\int \prod_{a=0}^{N^2-1} d^{d+1}L_a\;
\exp\bigg(iL_a\cdot(X_{2a}-X_{1a})-i\frac{L_a^2}{2Nm}(t_2-t_1)\bigg),
\\
&~&S_\fp [j+1,j;\Delta t]=\sum_{a=0}^{N^2-1}
\frac{Nm(X_{j+1,a}-X_{j,a})^2}{2\Delta t},
\eea
we find for (\ref{amplitude-1})
\bea
S_{FI}&\sim& \prod_{a=0}^{N^2-1} \delta^d(P_{Fa}-P_{Ia})
\delta(E_{Fa}-E_{Ia})
\nonumber\\&+& (\cdots) \int_{t_I}^{t_F}dt\int d^d X 
\int d^dX_I\; d^dX_F
\prod_{a=0}^{N^2-1}\bigg( \e^{i(E_{Fa}t_F-E_{Ia}t_I)}
\e^{-i(P_{Fa}\cdot X_{Fa}-P_{Ia}\cdot X_{Ia})}\bigg)
\nonumber\\&~&
\int \prod_{b=0}^{N^2-1} d^{d+1}Q_b\;
\exp\bigg(iQ_b\cdot(X_{Fb}-X_b)-i\frac{Q_b^2}{2Nm}(t_F-t)\bigg)
\nonumber\\&~&
\bigg(
\sum_{c=0}^{N^2-1} 
\bigg\{\exp \bigg(iq \epsilon\cdot \bigg(\frac{X_{Fc} - X_c}{t_F-t}+
\frac{X_c-X_{Ic}}{t-t_I}\bigg)\bigg) 
\bigg\}_{{\rm linear\;in}\;\epsilon}  
\tr\bigg(T^c \e^{ik\cdot X-i\omega t} \bigg)
\nonumber\\&~&
+\tr\bigg(iq\epsilon_0  \e^{ik\cdot X-i\omega t}\bigg) \bigg)
\nonumber\\&~&
\int \prod_{e=0}^{N^2-1} d^{d+1}L_e\;
\exp\bigg(iL_e\cdot(X_e-X_{Ie})-i\frac{L_e^2}{2Nm}(t-t_I)\bigg)
+O(q^2),
\eea
in which to make easier the calculation, we have exponentiated the
$\vec\epsilon$; so we should keep only the linear term in $\epsilon$
finally. In above the symbol $A\cdot B$ is for the inner product $A_iB^i$.
It is worth to recall that the spatial integrals like $\int d^d X$ are in
fact as $\int \prod_{a=0}^{N^2-1} d^d X_a$. Here we leave the term
$\epsilon_0$ for the reader. After doing the integrations on $d^dX_{I,F}$,
we have:
\bea
S_{FI}&\sim& \prod_{a=0}^{N^2-1} \delta^d(P_{Fa}-P_{Ia})
\delta(E_{Fa}-E_{Ia})+ (\cdots) \int_{t_I}^{t_F}dt\int  
\prod_{a=0}^{N^2-1} d^dX_a \e^{i(E_{Fa}t_F-E_{Ia}t_I)}
\nonumber\\&~&
\int \prod_{b=0}^{N^2-1} d^{d+1}Q_b\;d^{d+1}L_b\;
\e^{-i\frac{Q_b^2}{2Nm}(t_F-t)}
\e^{-i\frac{L_b^2}{2Nm}(t-t_I)}
\e^{i(L_b-Q_b)\cdot X_b}
\nonumber\\&~&
\bigg(
\sum_{c=0}^{N^2-1} 
\prod_{e=0}^{N^2-1}
\delta^d\bigg(Q_e-P_{Fe}+\frac{q\epsilon\delta_{ce}}{t_F-t}\bigg)
\delta^d\bigg(L_e-P_{Ie}+\frac{q\epsilon\delta_{ce}}{t-t_I}\bigg)
\nonumber\\&~&
\bigg\{\exp \bigg(iq \epsilon\cdot X_c 
\bigg(\frac{-1}{t_F-t}+\frac{1}{t-t_I}\bigg)\bigg) 
\bigg\}_{{\rm linear\;in}\;\epsilon}  
\tr\bigg(T^c \e^{ik\cdot X-i\omega t} \bigg)\bigg)
\nonumber\\&~&
+O(q^2).
\eea
By using the $\delta$-functions we simply can do the integrations on
$d^dQ$ and $d^dL$. Also, based on the fact that $\exp(ik\cdot
X)=\exp(ik\cdot X_\cm {\bf 1}_N) \exp(ik\cdot\hat{X})$, with
$\tr(\hat{X})= 0$, we can do the integration on $d^d X_0=d^d X_\cm$. By
recalling $E_a=P_a^2/(2Nm)$ for $I$ and $F$ states (convention
$\tr(T^aT^b)=N\delta^{ab}$), and by the limits $t_I\goes -\infty$
and $t_F\goes\infty$, we arrive at: 
\bea
S_{FI}&\sim&\prod_{a=0}^{N^2-1}\delta^d(P_{Fa}-P_{Ia})
\delta(E_{Fa}-E_{Ia})\nonumber\\
&+&(\cdots)
\delta^d (P_{F\cm}-P_{I\cm}-k)
\delta\bigg(\sum_{a=0}^{N^2-1}(E_{Fa}-E_{Ia})-\omega\bigg)
\nonumber\\&\times&
\int \prod_{b=1}^{N^2-1} d^d\hat{X}_b \;
\e^{i(P_{Ib}-P_{Fb})\cdot \hat{X}_b}
\sum_{c=0}^{N^2-1} 
\bigg\{\e^{\frac{i}{m}q \epsilon\cdot (P_{Fc}+P_{Ic})}
\bigg\}_{{\rm linear\;in}\;\epsilon}  \!\!\!
\;\tr\bigg(T^c \e^{ik\cdot \hat{X}} \bigg)+ O(q^2).
\eea


\end{document}